\documentclass{emulateapj}

\newcommand{\OIII}{\mbox{O\,\textsc{iii}}}

\newcommand{\OI}{\mbox{O\,\textsc{i}}}
\newcommand{\NII}{\mbox{N\,\textsc{ii}}}

\newcommand{\SII}{\mbox{S\,\textsc{ii}}}
\newcommand{\kms}{km sec$^{-1}$}

\shorttitle{A spatially-resolved spectroscopic study of a LINER galaxy}
\shortauthors{Bae et al.}

\begin{document}

\title{A Keck/LRIS spatially-resolved spectroscopic study of a LINER galaxy \\SDSS J091628.05+420818.7 }

\author{Hyun-Jin Bae\altaffilmark{1}, Jong-Hak Woo\altaffilmark{2*}, Masafumi Yagi\altaffilmark{3}, Suk-Jin Yoon\altaffilmark{1}, and Michitoshi Yoshida\altaffilmark{4}}

\altaffiltext{1}{Department of Astronomy and Center for Galaxy Evolution Research, Yonsei University, Seoul 120-749, Republic of Korea} 
\altaffiltext{2}{Astronomy Program, Department of Physics and Astronomy, Seoul National University, Seoul 151-742, Republic of Korea}
\altaffiltext{3}{Optical and Infrared Astronomy Division, National Astronomical Observatory of Japan, Tokyo 181-8588, Japan} 
\altaffiltext{4}{Hiroshima Astrophysical Science Center, Hiroshima University, Hiroshima 739-8526, Japan} 
\altaffiltext{*}{Corresponding author: woo@astro.snu.ac.kr}

\begin{abstract}
Using spatially-resolved spectra obtained with the Low Resolution Imaging Spectrometer 
at the Keck I telescope, we investigate the nature of ionizing sources and 
kinematic properties of emission-line gas in a LINER galaxy SDSS J091628.05+420818.7, 
which is a nearby ($z=0.0241$) and bright ($M_r=-20.2$) early-type galaxy.
After subtracting stellar absorption features using a combination of simple stellar 
population models, we measure the flux, line-of-sight velocity, and velocity dispersion 
of four emission lines, i.e., 
H$\alpha$, H$\beta$, [\OIII] $\lambda$5007, and [\NII] $\lambda$6584, 
to study radial change of emission-line fluxes and velocities.
Compared to the point-spread-function of the observation,
the emission-line region is slightly extended but comparable to the seeing size.
The central concentration of emission-line gas suggests that
ionization is triggered by a nuclear source,
excluding old stellar population as ionizing sources.
We find that emission-line gas is counter-rotating with respect to stellar component 
and that the [\OIII] $\lambda$5007 line is blueshifted compared to other 
emission lines, possibly due to an outflow.

\end{abstract}


\section{Introduction}

Active galactic nuclei (AGN) may play an important role in galaxy evolution 
by feeding energy to host 
galaxies, quenching star formation \citep[e.g.,][]{hop05,sp05} and leading to 
the black hole-galaxy scaling relations as observed in the local universe 
\citep{fm00,ge00,gu09,wo10}. 
By including simple AGN feedback prescriptions, recent semi-analytic models successfully
reproduced several observed properties including luminosity function, masses, and colors of galaxies \citep{cr06,bo06},  
although it is still unclear how AGN feedback works in quenching star formation. 
Observational studies have found that the majority of early-type galaxies exhibit LINER \citep[Low-Ionization Nuclear Emission-line Region,][]{he80} features at the central region \citep[e.g.,][]{ho97,an10,cb11}.
If LINERs are truly AGN, the high fraction of LINERs is consistent with the scenario
that the AGN feedback works as a star-formation quenching mechanism in early-type galaxies. 

In spite of the large population of LINERs in the local universe, the nature of ionizing source of LINERs is still debated.
On the one hand, LINERs have been generally regarded as a branch of low-luminosity AGNs \citep[e.g.][]{ho97,ka03,ke06}. This idea has been supported by several studies on nearby LINER galaxies, which provided a number of smoking-gun evidences for AGN-origin of LINER features, e.g, the presence of broad emission-lines \citep[e.g.,][]{ho97}, compact emission-line region \citep[e.g.][]{ke83,po00}, and ultra-violet (UV) flux variability \citep[e.g.,][]{maoz05}.

On the other hand, narrow-line features of LINER (i.e., Type 2 LINER) can be easily produced by other ionizing mechanisms since the LINER features are characterized by strong low-ionization lines, such as [\OI], [\NII], and [\SII], relative to high-ionization and the recombination lines. Several studies suggested possible mechanisms, e.g., evolved old stars, evolved
young stars, and supernovae shock, for producing the LINER-like features \citep{bi94,ds95,ta00}. 
By using photoionization models, for example, \cite{bi94} showed that evolved old stars, i.e., post-asymptotic giant branch (PAGB) stars, can produce Lyman continuum photons and exhibit LINER-like features, suggesting that a large fraction of LINERs 
among early-type galaxies can be originated by evolved stars \citep[see also][]{st08}. 
However, these models highly depend on the evolutionary tracks of PAGB stars, which are constructed under a number of theoretical assumptions \citep[e.g.,][]{sr00}. 
By comparing the measured line flux ratios of a sample of nearby galaxies 
with photoionization model predictions from \citet{bi94}, \citet{an10} reported that 11 out of 49 ($\sim 22$\%) LINER galaxies could be produced by PAGB stars, while the remaining $\sim 78$\% galaxies required additional ionization mechanisms. Since they adopt the model prediction of \citet{bi94}, one should again keep in mind the model uncertainties. 
In a similar way, \citet{ta00} suggested young evolved stars, i.e., planetary nebulae, from recent starburst as a viable mechanism for LINER-like features, implying a close connection between post-starburst and LINER-like phenomena. This scenario provides an alternative explanation of LINER features particularly for late-type galaxies.
In additoin to old and young stars, radiative shocks from oxygen-rich supernova 
can also produce LINER-like features without an AGN as suggested by \citet{ds95}. 

In contrast to the spatially concentrated LINER features originated from AGN \citep{ke83,po00},
those LINER-like emissions due to non-AGN mechanisms are tend to be spatially extended \citep{ve95,mon06,ri10,sa10}.
For example, \cite{po00} investigated the size of narrow-line region (NLR) of 14 LINER galaxies based on the high spatial resolution narrow-band imaging data from the \textit{Hubble Space Telescope} and showed that the most of the emission comes from central regions with the size of tens to hundreds of parsecs (pc), indicating that central AGN activities are responsible for the LINER features. 
In contrast, \citet{ri10} reported extended LINER emission in NGC 839, 
concluding that the LINER-like feature is related to other mechanisms than an AGN. 

It is also possible that LINER-like features observed with limited spatial resolutions may be a mixed bag of various ionization mechanisms \citep[e.g.,][]{fi96}. 
For nearby ($z < 0.01$) galaxies, spectroscopic observations generally include nuclear emission from the central $\sim 100$ pc region, avoiding possible contamination from an extended emission-line region. 
In contrast, it is far more complicated to understand the nature of emission lines in integrated spectra obtained with a large aperture; for example, Sloan Digital Sky Survey \citep[SDSS,][]{yo00}\footnote[1]{\url{http://www.sdss.org}} spectroscopic observations with a 3$\arcsec$-diameter fiber include emissions from much larger physical scale ($\sim 5$ kpc) in SDSS galaxies with a mean redshift $z \sim 0.1$. 
Hence, it is possible that even non-AGN galaxies can be interpreted as LINER galaxies due to contaminations from extended LINER-like emission region. 
While a number of studies classified SDSS LINERs as AGN \citep[e.g.,][]{ka03,ke06}, other studies argued that the SDSS LINERs may not be related to AGN activity \citep[e.g.,][]{st08,sa10,cf11,yb11}. 
For example, \cite{sa10} discussed that if SDSS LINER galaxies have modest equivalent width of [\OIII] $\lambda$5007 ($\lesssim 2.4$\AA), only a small fraction of these galaxies are truly powered by AGN. 
A recent statistical study by \citet{yb11} further claimed that AGN is not a primary ionizing source in the SDSS LINER galaxies although they cannot rule out the presence of a true AGN hidden in the integrated spectra obtained with a large aperture in SDSS.

Since the SDSS provides one of the largest galaxy sample in the local universe, revealing the true nature of LINERs in the SDSS sample is a crucial step in understanding AGN demography in the local universe, and the nature of the galaxy-AGN coevolution.
Spatially-resolved spectroscopy can shed light on the nature of ionizing source by constraining whether the emission is from nuclear or extended region.
The kinematic information of both stellar component and emission-line region can also provide useful hints on the dynamical relation between emission-line gas and stellar component.

To investigate the true nature of LINER features, we performed a pilot study using a carefully selected LINER early-type galaxy, SDSS J091628.05+420818.7, from SDSS DR7 \citep{ab09}.
In this work, we present detailed analysis on the emission-line properties based on the spatially-resolved spectroscopy.
The paper is composed as follows. Section 2 describes sample selection, observations and data reduction. 
Section 3 presents analysis method on the extracted spectra. We present the main results 
on the spatial profile and kinematics of emission lines in Section 4. 
Finally, we summarize and discuss our findings in Section 5. 
Throughout the paper, we use $\Lambda$CDM cosmological parameters, i.e., 
$H_{o}$ = 70 km s$^{-1}$ Mpc$^{-1}, \Omega _\Lambda = 0.73,$ and $\Omega _m = 0.27. $

\begin{figure}
\centering
\includegraphics[width=0.49\textwidth]{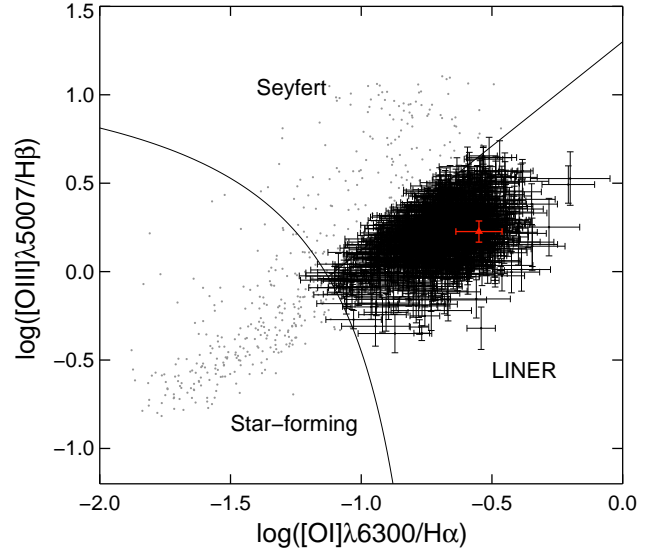}
\caption{The AGN diagnostic diagram for the LINER early-type galaxies selected from the SDSS. 
Gray dots represent 1382 nearby ($0.02 < z < 0.05$) and bright ($M_r < -19.7$) early-type galaxies with S/N $>$ 3.0 for four emission lines, i.e., H$\beta$, [\OIII] $\lambda$5007, [\OI] $\lambda$6300, and H$\alpha$. 
A sample of 902 LINERs (black dots with error bars) are indetified 
by adopting an empirical demarcation line of \cite{ke06} between Seyferts and LINERs.
We select a typical LINER early-type galaxies from the LINER sample (red point with error bar).
The curved line indicates a theoretical demarcation line between star-forming and AGN-host galaxies \citep{ke01}.}
\end{figure}

\begin{figure}
\centering
\includegraphics[width=0.44\textwidth]{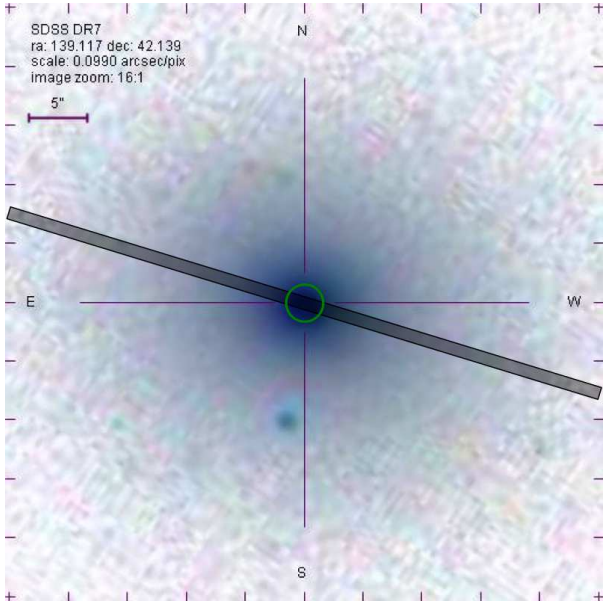}
\caption{SDSS optical ($gri$) composite color image of the selected LINER early-type galaxy, SDSS J091628.05+420818.7. The angle and the width of the black bar represent the position angle 
(+73$^\circ$) and the width (1\arcsec) of the slit, respectively. 
For comparison the 3$\arcsec$-diameter SDSS fiber size is shown at the center (green circle).}
\end{figure}

\section{Observations and Data Reduction}

\subsection{Sample Selection}

We selected a sample of low-redshift ($0.02 < z < 0.05$) and bright ($r$-band absolute magnitude $M_r < -19.7$) early-type galaxies with LINER features from SDSS DR7 to study the characteristics of LINER host galaxies (Bae et al. in preparation). 
To represent the general population of LINERs which are mostly hosted by early-type
galaxies and to avoid contribution from star-forming regions to the observed emission lines,
we first selected early-type galaxies based on $u-r$ color and $g-i$ color gradient using the criteria by \citet{pc05}.
Then, LINERs are identified following the criteria of \citet{ke06} on the emission-line ratios \citep[see Figure 1;][]{ba81}. 
We used four emission lines retrieved from the MPA-JHU catalog on the SDSS DR7 galaxies\footnote[2]{\url{http://www.mpa-garching.mpg.de/SDSS/DR7}}, i.e., H$\beta$, [\OIII] $\lambda$5007, [OI] $\lambda$6300, and H$\alpha$, with signal-to-noise ratio (S/N) larger than 3.0. 
To examine whether recent star formation can account for LINER features,
we further selected galaxies with the near-UV (NUV)$-$r color less than 5.5, which is an empirical demarcation line for the early-type galaxies with a recent ($\lesssim 1$ Gyr) star formation \citep{ka07}.

For a pilot study to constrain the nature of LINER emissions detected in SDSS galaxies, we chose from the sample one galaxy, SDSS J091628.05+420818.7, which represents general characteristics of the LINER early-type galaxies, with $M_r$ = $-$20.2 and the effective radius of 2.7 kpc. 
At the redshift of 0.0241, the galaxy provides a reasonable spatial resolution, 0.51 kpc arcsec$^{-1}$. Figure 1 shows emission-line properties of the sample of SDSS LINER early-type galaxies with $M_r < -19.7$ and $0.02 < z < 0.05$, as well as the selected galaxy.

The galaxy's morphology in the SDSS optical images indicates lack of recent and/or ongoing merger event (see Figure 2). 
Spatially extended UV-excess with NUV$-$r color was measured as 4.9 
in the NUV images from \textit{GALEX} \citep[Galaxy Evolution Explorer,][]{ma05}\footnote[3]{\url{http://galex.stsci.edu}}, suggesting that an existence of spatially extended young stellar population. 
No radio counterpart for the galaxy is detected at 1.4 GHz through the VLA FIRST survey \citep{be95}\footnote[4]{\url{http://sundog.stsci.edu}}. 

\subsection{Observations}

We performed a spatially-resolved spectroscopy on the selected target, SDSS J091628.05+420818.7, using the Low Resolution Imaging Spectrometer \citep[LRIS,][]{ok95, ro10} at the Keck I telescope with 1 hour (6 $\times$ 600 seconds) exposure on 31 March 2011. 
The D560 dichroic mirror was used to split the incoming light into the red and blue side CCDs, and the 600/4000 grism and the 600/7500 grating were used for the blue and red sides, respectively.
The instrumental setup was chosen to cover the spectral range from 3200\AA\ to 9000\AA, including forbidden lines [\OIII] $\lambda$5007, [\NII] $\lambda$6584, and Balmer lines, i.e., H$\alpha$, H$\beta$, which are essential ingredients to identify LINERs.
We used a 1$\arcsec$-wide slit with the position angle of +73$^\circ$ counter-clockwise direction from the North to the East, which was determined as a major axis based on  $IRAF$/{\textsf ellipse} fitting using the SDSS $g$-band image. 
To increase S/N, we used 2$\times$2 binning in the CCD, resulting in a spatial scale 0.$\arcsec$27 pixel$^{-1}$.
The dispersion scales for the blue and red sides are 1.23\AA\ pixel$^{-1}$ and 1.60\AA\ pixel$^{-1}$, respectively.

To calibrate the flux and estimate the seeing size, we observed two spectrophotometric stars, Feige 34 and Feige 66 \citep{ok90}, 
right before and after observing the target galaxy.
The seeing size decreased during our observations from $\sim 1.\arcsec6$ to $\sim 1.\arcsec1$ and from $\sim 1.\arcsec2$ to $\sim 0.\arcsec9$ on the blue and red side, respectively. 
The effect of seeing and its variation is discussed in Section 3.3.

\begin{figure*}
\centering
\includegraphics[width=0.79\textwidth,angle=90]{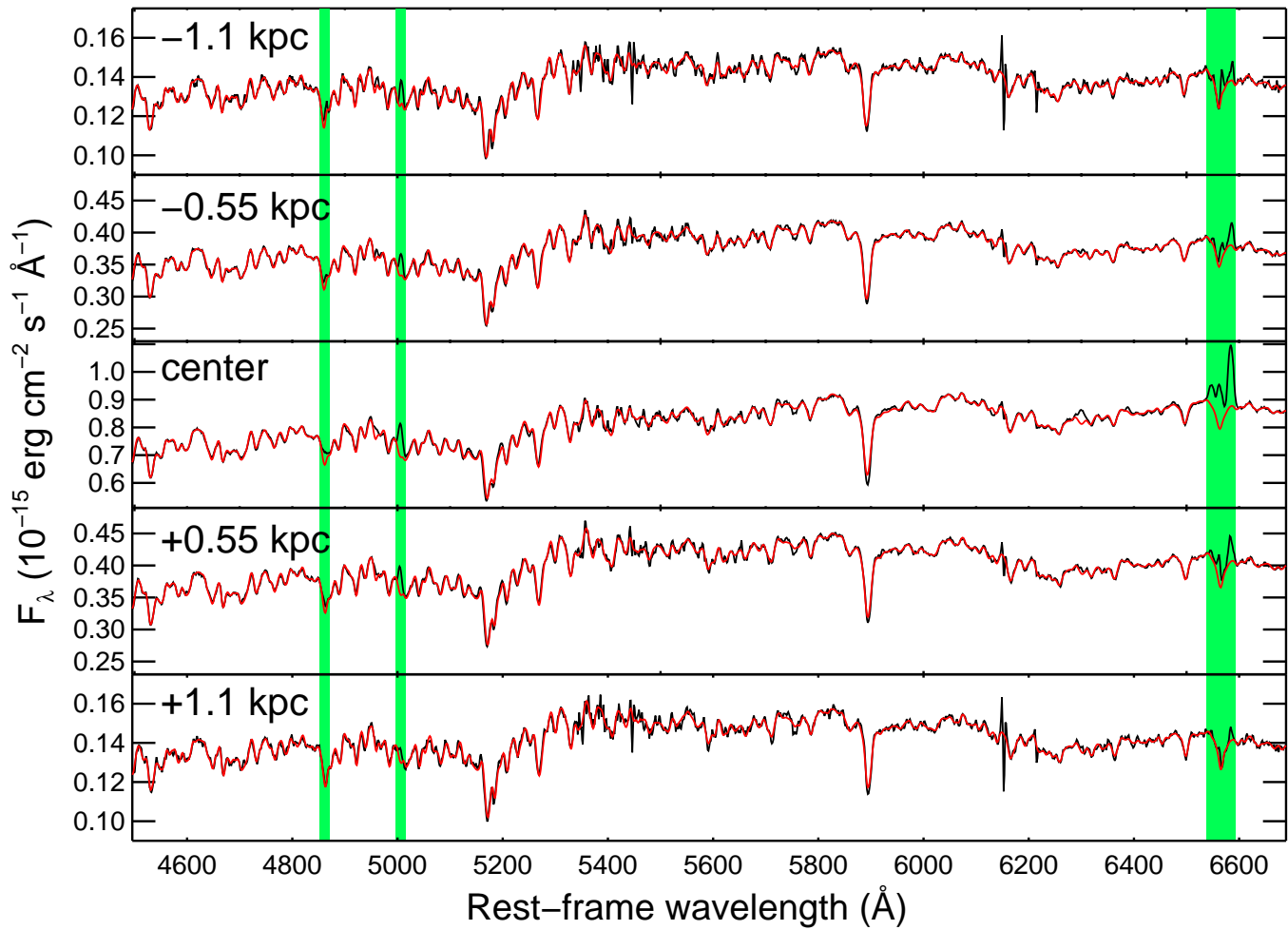}
\caption{The observed spectra at each apeture along the slit (black line) are plotted over the best-fit stellar models (red line) at each panel. The corresponding distance from the center is denoted in the upper-left conner. 
The spectral range of each emission line, i.e., H$\beta$, [\OIII] $\lambda$5007, and H$\alpha$+[\NII] doublet, is makred with green shades from shorter to longer wavelength, respectively. All spectra are shifted to the rest frame using the velocity measured from 
the central aperture.}
\end{figure*} 

\subsection{Data reduction}

We reduced the spectroscopic data using a series of $IRAF$ scripts including $twodspec$ package as previously performed \citep[e.g.,][]{wo06, mcgill08}.
The reduction process includes standard procedures, i.e., bias subtraction, flat-fielding, wavelength and flux calibrations.
After bias subtraction and flat-fielding, we removed cosmic rays using the Laplacian cosmic ray identification algorithm \citep{va01}. 
Arc lines (Hg, Ne, Ar, Cd, and Zn) and sky emission lines were used to calibrate wavelength scales for blue and red sides, respectively. 
Atmospheric extinction was corrected for each exposure while the galactic extinction was corrected using \textsf{deredden} task.
Flux-calibration was performed using two spectrophotometric standard stars Feige 34 and Feige 66, which were observed with the same instrumental setup during the observing run.
After subtracting sky background by adopting sky regions at $>  20$\arcsec\ from the galaxy center, we extracted nine spectra consecutively along the spatial direction using a 4-pixel aperture diameter, which is equivalent to 1.$\arcsec$08 and $\sim 0.55$ kpc at the distance of the target. However, we only used five central apertures in our analysis, since we could not detect weak emission lines, especially H$\beta$, from the spectra extracted from the outer region. 
Thus, we cover the spatial range from $-$1.1 kpc (SW) to +1.1 kpc (NE) with respect to the center of the galaxy. 

Finally, we combined the extracted spectra from each exposure with a median filtering, after scaling fluxes to a mean flux averaged over from 4950\AA\ to 5150\AA\ and from 6400\AA\ to 6800\AA\ for blue and red sides, respectively. 
The selected ranges are most preferred since they include useful emission lines for AGN diagnostics, e.g., H$\beta$ and [\OIII] $\lambda$5007 on the blue side, [OI] $\lambda$6300, H$\alpha$, and [\NII] $\lambda$6584 on the red side. 
Using arc and sky emission lines, we measured the instrumental resolutions as 6.1\AA\ and 4.2\AA\ full width at half maximum (FWHM), respectively for blue and red sides, corresponding to Gaussian velocities $\sim 152$ and $\sim 80$ km s$^{-1}$ at the wavelength of the observed H$\beta$ and H$\alpha$.
The final reduced spectra along the spatial direction are presented in Figure 3.

\section{Analysis}

In this section, we describe the method for measuring the emission-line fluxes.
Since it is crucial to remove stellar absorption features to reliably measure the emission line fluxes, we first present the procedure for subtracting stellar absorption line \citep[e.g.,][]{ho08}. 
Then, we present the emission line flux measurements. 
Lastly, we describe how we construct the point-spread-function (PSF).

\subsection{Stellar Continuum Subtraction}

In order to subtract the stellar absorption lines, we need to construct a best-fit stellar model with a combination of stellar templates or population models. Here we exploit 47 MILES simple stellar population (SSP) templates \citep{sb06}, which include the population age from 60 Myrs to 12.6 Gyrs, with twice-solar metallicity ($Z=0.03$) and the Kroupa initial mass function \citep{kr01}. For a consistency check we compare the results using simple stellar population models with solar metallicity ($Z=0.02$), leading to little difference with slightly higher $\chi^2$ value of the fitting. 

We construct a best-fit stellar model for galaxy spectra using the penalized pixel-fitting code \citep[pPXF,][]{ce04}\footnote[5]{http://www-astro.physics.ox.ac.uk/$\sim$mxc/idl}, which is widely used in stellar continuum subtraction and gas-kinematics analysis for individual galaxies and large surveys \citep[e.g..][]{fb04,ca11}. 
The pPXF implemented the MPFIT code \citep{ma09}, which is based on the Levenberg-Marquardt algorithm \citep{mo78} for $\chi ^2$ minimization.
For the fitting process, we use the limited rest-frame wavelength range 
from 3550\AA\ to 7380\AA, to match the available wavelength range of the 
MILES SSP templates
  
We mask out the wavelength ranges of well-known nebular emission lines and sky absorption 
features during the fitting process.
To account for imperfect flux calibration and increase the quality of the fit,
we adopt a multiplicative polynomial function. 
Note that we choose a high order (100-200) polynomial 
depending on the shape of spectra, however the polynomial function does not 
significantly affect the results since the polynomial function does not 
attempt to fit any single emission line.
During this process, we simultaneously measure the first (line-of-sight velocity)
and second moments (velocity dispersion) of the stellar absorption lines
using spectra obtained at each aperture, which are used to 
construct the radial profile of rotational velocity and velocity dispersion 
of stellar component as described in Section 4.2.
To measure stellar velocity dispersion, we use the $\sim 200$\AA\ range of spectra 
including the Mg\textit{b} triplet and Fe (5270, 5300\AA) lines.
We correct for the instrumental resolutions of the observed spectra and 
the MILES SSP templates (FWHM=2.51\AA) by subtracting them in quadrature (see Eq. 1 in
Woo et al. 2004). The typical measurement errors of stellar velocity dispersion 
determined from the fitting procedure are $\sim 5$\%.
Using the spectrum extracted from the centeral aperture, 
we measure stellar velosity dispersion as $178 \pm 10$ km s$^{-1}$, 
which is consistent with $182 \pm 3.8$ \kms\ measured from the 3$\arcsec$ diameter fiber
spectrum of SDSS.
In Figure 3, we present the best-fit stellar models overlaid with 
the observed spectra at each aperture. All spectra are shifted from the observed-frame 
to the rest-frame using the radial velocity measured from the central aperture. 

After the continuum fit, we subtract the stellar model from the observed spectra, 
revealing the stellar absorption corrected nebular emission lines, 
which are used in emission-line measurement (see Section 3.2 and Figure 4).
The standard deviations of the residuals from stellar subtraction 
(excluding emission-line regions) are $7.0\times 10^{-18}$ erg cm$^{-2}$ s$^{-1}$ \AA$^{-1}$ and $1.8\times 10^{-18}$ erg cm$^{-2}$ s$^{-1}$ \AA$^{-1}$ for apertures at the center and outer (+1.1 kpc), respectively.
For comparison, peaks of flux density for each emission line are significantly larger than the residual from stellar subtraction ($>3\sigma$), 
confirming that the stellar fit is good enough to measure emission-line fluxes.
In case of the H$\beta$ at +1.1 kpc aperture, which is the weakest emission-line we measured, the peak of flux density is comparable to the fitting residual, indicating that it is difficult to measure the weak H$\beta$ emission-line flux at +1.1 kpc aperture.

\begin{figure}
\centering
\includegraphics[width=0.54\textwidth]{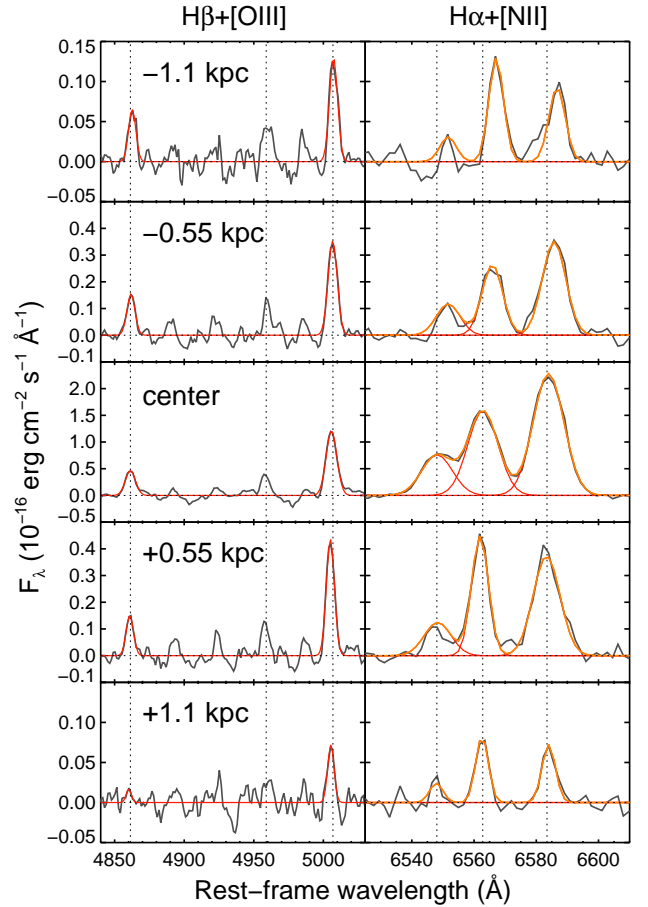}
\caption{Residual emission lines (gray line) after subtracting stellar continuum
and the best-fit emission line model (red line) for the H$\beta$+[\OIII] (left panel) 
and the H$\alpha$+[\NII] regions (right panel), repectively. 
Vertical lines represent the reference wavelength of each emission line.
The distance from the center is denoted in left panels nd side. 
Orange lines on the right-hand side panels exhibit the sum of three Gaussian functions for H$\alpha$ and [\NII] doublet.} 
\end{figure}

\subsection{Emission-line Measurement}
After the stellar subtraction, we measure four emission lines for each aperture; H$\beta$, [\OIII] $\lambda$5007, H$\alpha$, and [\NII] $\lambda$6584. The wavelength range we used also contains the other emission lines, such as [\SII], [\OI], that are typically used in AGN diagnostics. However, we could not use those lines since the [\SII] $\lambda\lambda$ 6716, 6731 lines fall in the atmospheric absorption B band and the [\OI] $\lambda$6300 line is too weak to properly measure line flux. In Figure 4, we present the emission lines after subtracting stellar component and the best-fit single Gaussian models.

We use the Monte-Carlo technique to measure the mean strength, wavelength, dispersion and estimate their uncertainties for each emission line simultaneously. 
First of all, we take a standard deviation of the continuum fitting residuals within $\sim \pm 100$\AA\ range from the center of each emission line as a proxy for systematic uncertainties, which is larger than the measurement 
uncertainties in the emission-line fitting process. Thus the uncertainties 
adopted in the following simulations should be considered as upper limits. 
Then, we generate a thousand simulated spectra using the random values at each pixel, which follow a normal distribution 
with the systematic uncertainty as a standard deviation, and add them to the emission lines. 
For each simulated spectrum, we measure line fluxes using a Gaussian model fitting method as follows. 
For the H$\alpha$+[\NII] region, we use triple Gaussian models by constraining the [\NII] $\lambda$6584-to-[\NII] $\lambda$6548 ratio as 3.0, while we use a single Gaussian model for other emission lines.
In this procedure, we measure the central wavelength and line dispersion and their uncertainties of each emission line simultaneously, in order to investigate the kinematics of nebular emission. 
We correct for the instrumental broadening by subtracting the instrumental resolution from the measured line dispersion in quadrature as generally practiced \citep[e.g.,][]{ba02,wo04}. 
In case of the [\OIII] $\lambda$5007 at +1.1 kpc aperture, the measured line dispersion ($\sim 2.4$\AA) is comparable to the instrumental resolution ($\sim 2.5$\AA\ for the blue side), hence we did not correct for the
instrumental resolution and the measured line dispersion of [\OIII]
should be taken as an upper limit.
 
Systematic uncertainty of the radial velocity comes from the uncertainty of wavelength 
calibration, which was estimated as$\sim 0.2$\AA\ by adopting the standard deviation 
of arc or sky line wavelengths with respect to the wavelength solution. The measurement 
errors of the line velocity derived from the Monte-Carlo realization are less than or 
comparable to the systematic error in most cases. Hence, we added both errors in quadrature 
to obtain the final uncertainty of each emission-line velocity.
For the line dispersion measurements, on the other hand, we only consider the measurement error derived from the Monte-Carlo realization. 

\begin{figure}
\centering
\includegraphics[width=0.495\textwidth]{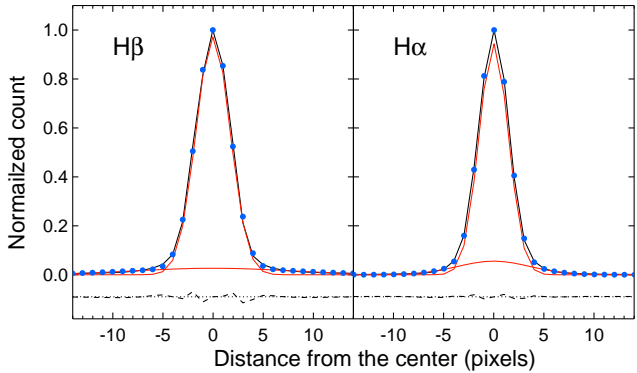}
\caption{PSF models based on the observation of Feige 66 
for H$\beta$ (blue side; left panel), and H$\alpha$ (red side; right panel). 
The photon counts measured at each pixel along the slit are normalized 
by the peak value (blue dots). 
Two Gaussin functions (red lines) are used for the best-fit PSF model
(black line). 
The residuals are shown at the bottom of each panel (dashed line).}
\end{figure}

\subsection{Point Spread Function}
A precise measurement of a PSF is an important task to examine whether the LINER-like features are originated from an extended or nuclear region.
To measure the PSF we utilize the spectrophotometric stars observed right before
and after observing the target galaxy. Using photon counts of each pixel from standard star observations, 
we obtained the spatial profile of PSF at each wavelength range which is corresponding to the spectral range of each emission lines. For each emission-line region, we performed a median combine by using 10 pixels along the dispersion direction, then we constructed a model PSF by fitting the observed PSF with multi-Gaussian functions,
resulting in one PSF for each emission line. 

Figure 5 illustrates the PSF model based on the spatial distribution of a 
standard star Feige 66.  Note that the spatial distribution is slightly extended
than a single Gaussian. Thus, we used two Gaussian components for the best PSF model. 
In the H$\beta$ region (Fig. 5 left), 
we used FWHM 4.$\arcsec$91 and 1.$\arcsec$08, respectively for
weak and strong Gaussian compoents with the peak ratio of weak to strong components as 0.03 
while in the H$\alpha$ region (Fig. 5 right) FWHM was 4.$\arcsec$91 and 1.$\arcsec$08, respectivley for weak and strong Gaussian components with the peak ratio of 0.06.


\section{Results}
In this section, we present the results of spatial profiles of emission-line flux and kinematic analysis for emission-line gas and stellar component, and suggest the most probable ionizing sources for the observed LINER features, and possible explanations for the observed kinematics.

\subsection{Spatial Profiles of Emission-line Flux}

\begin{figure}
\centering
\includegraphics[width=0.45\textwidth,angle=90]{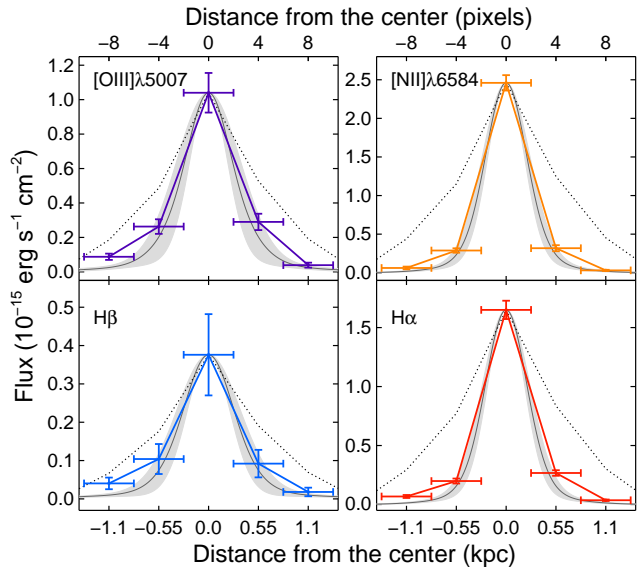}
\caption{Spatial profiles of emission-line flux: [\OIII] $\lambda$5007 (top left), 
H$\beta$ (bottom left), [\NII] $\lambda$6584 (top right), and H$\alpha$ (bottom right). 
Thick solid lines indicate the profile of each emission line, and vertical error bars indicate a standard deviation of the measured fluxes from the Monte-Carlo realization, while horizontal error bars represent the size of aperture for each emission-line flux. Dotted lines are normalized stellar continuum flux at 5100\AA\ ($F_{5100}$). Gray shade for each panel indicates that the minimum-to-maximum PSF range of the observation. We also plot the mean of both PSFs along the slit with a thin solid line over the gray shade.}
\end{figure}

To examine whether the emission-line region is spatially extended, 
we investigate the radial distribution of each emission line by comparing 
the spatial profiles of emission lines with that of stellar component
and the PSF determined in Section 3.3.
Figure 6 presents the spatial profiles of the emission-line flux for  [\OIII] $\lambda$5007, H$\beta$, [\NII] $\lambda$6584, and H$\alpha$ along the slit.
We overlay the PSF determined in Section 3.3, after normalizing it 
to the emission-line flux at the center. 
Since the exact PSF at the time of exposure on the galaxy is not known, 
we assume the true PSF size is between the maximum and minimum PSF sizes 
of spectrophotometric stars as determined in Section 3.3, as
plotted with black solid lines and gray shades. 

We find that the spatial profiles of emission-line flux are slightly more extended than,
but comparable to the PSFs, indicating that the emission-line gas is centrally
concentrated. At $-$1.1 kpc, there is a noticeable excess of emission-line fluxes 
compared to the PSF, suggesting that the emission-line region is slightly more 
extended than a point source. 
We compare the spatial profiles of emission-line region and stellar component,
to investigate a physical link between stars and ionizing sources of the emission lines.
In Figure 6, the spatial profile of stellar continuum flux at 5100\AA\ ($F_{5100}$) 
is presented after normalizing it to each emission-line flux at the center. 
Compared to the spatial distribution of stellar component, the emission-line
region is centrally concentrated, implying that old stellar population may not
be a primary ionizing source of the LINER feature.
Unless old stellar population is concentrated at the galaxy center, 
there seems no direct physical link between the emission-line gas and old stars. 

We note that the spatial profiles of the emission-line flux are asymmetric, while that of 
stellar continuum flux is symmetric.
In Figure 6, for example, the line fluxes of [\OIII] $\lambda$5007 and H$\beta$ at $-$1.1 kpc aperture are 
a factor of two larger than those of the same emission lines at $+$1.1 kpc aperture
(see also Figure 4).
Such asymmetric spatial profiles of the emission lines further suggest that 
ionizing source may not be not physically linked to old stellar component.

We investigate whether the distribution of young stellar population is 
asymmetric by measuring the 4000\AA\ break index $D_n$(4000) as defined by \cite{ba99} since  $D_n$(4000) is an indicator of the presence of young stars.
The $D_n$(4000) indices measured from -1.1 kpc to 1.1 kpc apertures are, 
respectively, [1.81, 1.86, 1.95, 1.87, 1.80] with measurement error 
less than 0.01, showing that old stellar population is dominant for all aperture but with a frost of young stellar population. 
The symmetric trend of the $D_n$(4000) indicates that the distribution 
of young stellar population is also symmetric, 
further implying that ionizing source is not primarily linked to both old and young stellar populations. 

\begin{figure}
\centering
\includegraphics[width=0.48\textwidth]{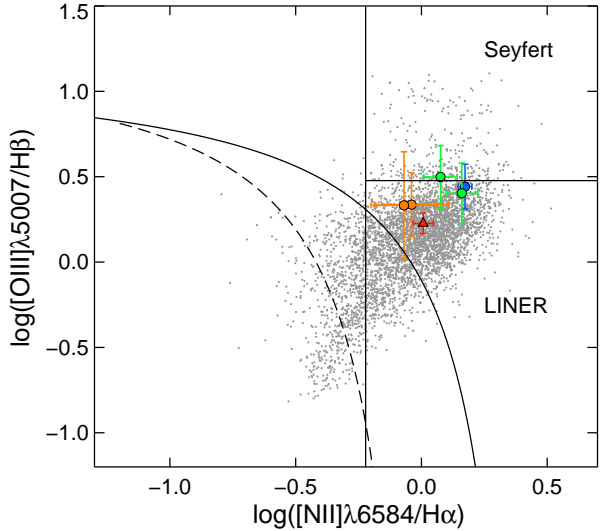}
\caption{The AGN diagnostics based on the emission-line ratios for each aperture. 
Compared to the emission-line ratio obtained from the SDSS integrated spectrum (red dot),
the emission-line ratios measured from spatially resolved spectra are shown 
for the galaxy center (blue), $\pm 0.55$ kpc (green) and $\pm 1.1$ kpc (orange), 
respectively. 
SDSS early-type galaxies described in Section 2.1 are plotted for comparison
(grey dots). Dashed and solid curved-lines indicate empirical and theoretical demarcation lines between star-forming and AGN-host galaxies, respectively \citep{ka03,ke01}. Horizontal and vertical solid lines are demarcation lines for Seyfert and LINER galaxies, indicating [\OIII] $\lambda$5007/H$\beta$ = 3.0 and [\NII] $\lambda$6584/H$\alpha$ = 0.6, respectively\citep{ka03}.
}
\end{figure}

Figure 7 presents the AGN diagnostic diagram with flux ratios of [\NII] $\lambda$6584 to H$\alpha$ and [\OIII] $\lambda$5007 to H$\beta$ measured at each aperture. We find that 
both line ratios tend to decrease from the central to outer ($\pm 1.1$ kpc) apertures,
implying that ionizing power varies as a function of the radial distance from the center.
However, we note that the decrease of the [\NII]/H$\alpha$ flux ratio can be 
interpreted as the decrease of gas density and/or metallicity.
To compare with the flux ratio measured from fiber observations, 
we plot the fiber-integrated emission-line ratios based on the SDSS spectrum. 
The integrated emission-line ratios are $\sim 0.2$ dex lower than those measured at the center, implying that the emission-line ratios based on the fiber-integrated spectra can be significantly underestimated compared to the central emission-line ratios.

\begin{figure}
\centering
\includegraphics[width=0.48\textwidth]{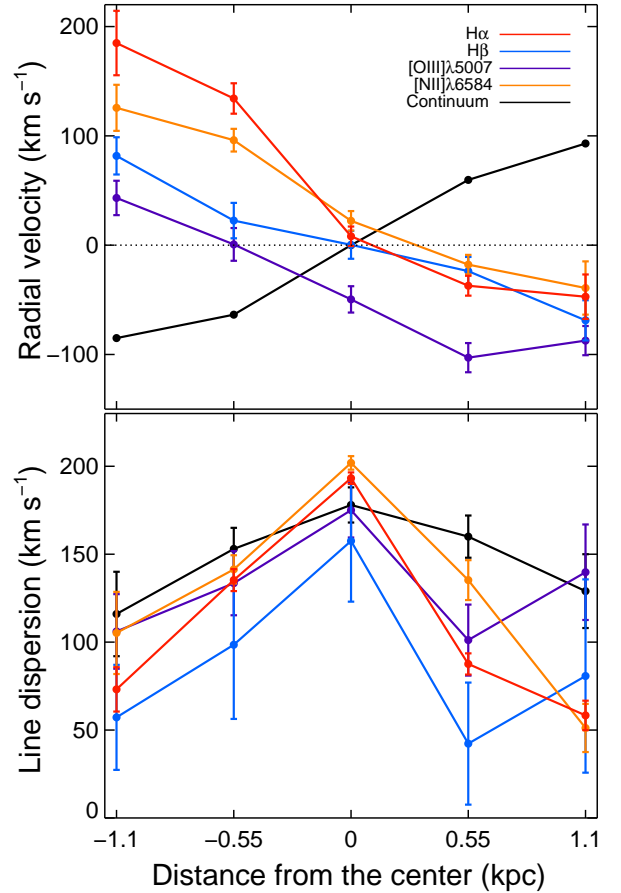}
\caption{Kinematic properties of the nebular emission and stellar continuum along the slit. The upper and bottom panels show the spatial profiles of the line-of-sight velocity 
and velocity dispersion, respectively. Adopted color codes are the same as in Figure 6. 
The uncertainties of the line velocities are estimated by adding systematic and measurement 
errors in quadrature (upper panel) while the error bars of line dispersion indicate
the measurement errors (lower panel).} 
\end{figure}

\subsection{Kinematics of Emission-line Gas}

Comparing the kinematic properties of emission-line gas and stellar component can 
provide an important clue on the nature of the ionizing sources. 
In Figure 8 we present the line-of-sight velocities and velocity dispersions of each 
emission line and stellar component measured at five different locations along the slit. 
The line-of-sight velocity measured from stellar lines at the galaxy center is $7218 \pm 2$ km s$^{-1}$, which is consistent within the error with the SDSS redshift $z=0.0241 \pm 0.0002$, corresponding to $7225 \pm 60$ km s$^{-1}$. 
By adopting the line-of-sight velocity measured from stellar component 
at the galaxy center as a zero-point, we present the relative velocities 
of gas and stellar components. 
For emission lines, we adopt the air wavelength provided by \cite{mo91} 
as a rest-frame wavelength to calculate the radial velocity of each emission line.

We find several characteristics of the emission-line gas and stellar component. 
First, there is a rotating stellar component with a line-of-sight velocity of $\sim \pm 90$ km s$^{-1}$ at $\sim \pm 1$ kpc. Since the old stars are dominant in the luminosity-weighted 
galaxy spectra, the rotation is mainly represented by old stellar population. 
It is not unusual to find a rotating stellar component among early-type galaxies \citep[e.g.][]{em07,em11}. 
For example, based on the integral field spectroscopic observations 
of a sample of 48 nearby early-type galaxies, \citet{em07} reported that
all early-type galaxies in their sample were either slow or fast rotator. 

Second, emission-line gas is counter-rotating with respect to stellar component, 
indicating that gas and stellar component are kinematically decoupled. 
The dynamical detachment implies that emission-line region may not be directly 
related to stellar component although we cannot rule out the stellar component
as ionizing sources solely based on the dynamical decoupling. 

Third, [\OIII] $\lambda$5007 emission line is significantly blueshifted 
with respect to stellar component, by $-56 \pm 12$ km s$^{-1}$ at the center, 
while other emission lines do not show clear velocity offset. 
An artifact due to a misaligned slit position may cause the velocity offset
\citep{co09}, however, the detected offset is not an artifact 
since the radial velocities of all other emission lines except 
[\OIII] $\lambda$5007 are consistent with the reference velocity of stellar lines.

While the velocity offset of emission-line gas at the center of LINER galaxies has not been reported\footnote[6]{Although \citet{ba08} reported that a radial velocity offset in one of LINER nuclei in NGC 3341, it is due to remote LINER nucleus from the center.}, 
it is rather common in Seyfert galaxies. 
Investigating [\OIII] $\lambda$5007 line for 65 local Seyfert galaxies, for example, \cite{cr10} summarized 
that 6\% of the Seyferts display redshifted velocity offset with $\Delta v \ge +50$ km s$^{-1}$, and 35\% display blueshifted offset with $\Delta v \le -50$ km s$^{-1}$. They provided a physical interpretation that the offset is due to the combined effect of gas outflow of ionized gas in the NLR and dust extinction in the inner stellar disk \citep[see Figure 7 of ][]{cr10}.           
Although this scenario was developed for Seyfert galaxies, it is a viable 
solution for the velocity offset detected in the LINER galaxy,
assuming that as a minor version of Seyfert galaxies, LINER galaxies also have 
mass outflow at the center \citep[e.g.,][]{ke06}. 
Compared to other scenarios, e.g., infalling small galaxy to the center of an AGN-host galaxy \citep{he09} or alignment of two AGNs \citep{sh09}, the outflow scenario provides a simpler 
explanation for the velocity offset of emission-line gas.
Following the outflow scenario, it is probable that we only detected the [\OIII] velocity offset because [\OIII] ionization region is closer to the accretion
disk, hence, more affected by an outflow \citep[e.g.,][]{ko08}.

Counter-rotating gas disks are often found in nearby early-type galaxies based on longslit spectroscopy, e.g., NGC 4621 \citep{be90}, NGC 4450 \citep{ru92}, and early-type galaxies of recent IFU surveys \citep[e.g.][]{em07,em11}. 
The observed velocity offset of [\OIII] may be explained by that 
outflow gas is mixed with the counter-rotating disk,
producing the velocity offset with respect to stellar component
and the rotation at the same time. 
It seems that spatial distribution and dynamics of NLR gas in the LINER galaxy is rather 
complex as similarly found in other local LINER galaxies \citep{po00}. 

We investigate whether the LINER has an appropriate accretion rate to generate an outflow. Based on the empirical relationship between the mass of supermassive black hole ($M_{\rm BH}$) and stellar velocity dispersion \citep{gu09}, we estimate black hole mass as $\approx 1.1\times10^8$ $M_{\odot}$ from the stellar velocity dispersion $\sim 178 \pm 10$ km s$^{-1}$ measured at the galaxy center. 
To measure bolometric luminosity ($L_{bol}$) we use the narrow H$\alpha$ luminosity. Although it is more difficult to measure the $L_{bol}$ of Type 2 AGNs than that of Type 1 AGNs 
due to the lack of broad emission-lines, the narrow H$\alpha$ line can be used as a proxy for the broad emission-lines and the bolometric luminosity \citep{ho09a}. By adopting the bolometric correction factor $C_{\rm H\alpha} \sim 100$ \citep{ho09a}, we obtain $L_{bol} \approx  6.0\times 10^{40}$ erg s$^{-1}$ from the narrow H$\alpha$ line luminosity at the galaxy center. Combining the $M_{\rm BH}$ and $L_{bol}$ estimates, we derive the Eddington ratio as $\sim 4.2 \times 10^{-6}$, which is similar to the Eddington ratios of nearby LINER galaxies \citep{ho09a}.
In such a low Eddington ratio regime, AGN outflow is unlikely to be present due to the low radiative pressure compared to the gravitational force \citep[e.g.,][]{ra10}.

In contrast to the theoretical expectations, however, several observational studies reported LINER galaxies with a distinguishable outflow \citep[e.g.,][]{wa08,ma11}. For example, \citet{wa08} presented a clear evidence for nuclear outflow of several nearby LINER galaxies. In addition, NLR morphologies of nearby LINER galaxies are disturbed \citep{po00}, implying that the disturbed morphology is closely related to the outflow from the center \citep{ma11}. 
Although it is still puzzling that how LINERs with low Eddington ratios can exhibit an outflow, the observed kinematics of the emission-line gas in the LINER galaxy seems to be connected with an outflow from the central AGN.

\section{Discussions and Summary}

As a pilot study for investigating the nature of ionizing sources in LINERs,
we observed a LINER early-type galaxy, SDSS J091628.05+420818.7 with Keck/LRIS to obtain
spatially resolved spectra.
By comparing with the PSF of the observations, we find that the emission-line region 
is slightly extend, but comparable to the seeing size, indicating that
emission-line gas is centrally concentrated and triggered by a central source. 
We find that emission-line gas is counter-rotating with respect to stellar component and a significant radial velocity offset of [\OIII] $\lambda$5007 line.  

We suggest a scenario that gas outflow in the NLR is ionized by a central source, 
presumably an active black hole, and mixed with gas in the counter-rotating disk.
Though it is still difficult to disentangle the gas origin in this work, the gas is unlikely to be ionized and originated from the evolved stars from the old stellar population, since the gas is kinematically decoupled with the old stellar population. 
The counter-rotating gas may have external \citep[e.g.][]{ru92}, or internal origins such as evolved stars \citep[e.g.][]{ho09b}, and the gas may be causally linked with the existence of young stellar population. 
We cannot exclude the possibility that the young stars could be an ionizing source, but the contribution would be much weaker compared to that of AGN as the spatially concentrated profiles of emission-line flux indicate.  

Although some LINER galaxies may show an extended emission-line region, 
it is still possible that an active black hole is hidden in the integrated
spectra. Thus, we cannot simply exclude the LINER galaxies identified in SDSS
from the AGN population in the local universe. 
Detailed studies based on spatially-resolved spectra are necessary to reveal 
the true nature of the ionizing source of LINERs.

\acknowledgments

We thank Hai Fu and Luis C. Ho for their constructive suggestions.
The work of HJB was supported by the project of National Junior Research Fellowship which National Research Foundation of Korea conducts from 2010 and Hi Seoul Science Fellowship from Seoul Scholarship Foundation. JHW acknowledges support by Basic Science Research Program through the National Research Foundation of Korea funded by the Ministry of Education, Science and Technology (2010-0021558).
This study was partly supported by KAKENHI 21540247. The data presented herein were obtained at the W.M. Keck Observatory, which is operated as a scientific partnership among the California Institute of Technology, the University of California and the National Aeronautics and Space Administration. The Observatory was made possible by the generous financial support of the W.M. Keck Foundation. The authors wish to recognize and acknowledge the very significant cultural role and reverence that the summit of Mauna Kea has always had within the indigenous Hawaiian community. We are most fortunate to have the opportunity to conduct observations from this mountain. The observation was supported by Subaru Telescope which is operated by the National Astronomical Observatory of Japan, as a Subaru/Keck time exchange program.

\bibliographystyle{apj}


\clearpage

\end{document}